\documentclass{mem}
\usepackage{natbib}
\usepackage{txfonts}\usepackage{balance}
\usepackage{graphicx}
\usepackage[a4paper,breaklinks,dvipdfm]{hyperref}
\idline{91}{105}
\begin{document}
\def\teff{$T\rm_{eff }$}
\def\kms{$\mathrm {km s}^{-1}$}

\title{
How stars in globular clusters reveal the depletion of the Spite plateau of lithium}

   \subtitle{}

\author{
A.J.~Korn
          }

\institute{
Observational Astrophysics, Department of Physics and Astronomy, Uppsala University, Uppsala, Sweden\\
\email{andreas.korn@physics.uu.se}
}

\authorrunning{Korn }

\titlerunning{Lithium in GCs}

\abstract{I summarize the results of 15 years of research on the surface abundances of stars in globular clusters. Globular-cluster stars afford a unique view of the physical processes that shape the surface abundances of stars on and beyond the Spite plateau, as one can study stars in different evolutionary phases with the same composition at birth. The main result is the finding of observational trends in surface abundances of heavy elements (Mg, Ca, Ti, Cr, Fe, Ba) that range from $\simeq$0.1\,dex (M4 at [Fe/H]=$-1.1$) to $\simeq$0.3\,dex (M30 at [Fe/H]=$-2.3$) with turn-off and subgiant stars generally showing lower abundances than giant stars. 
Any model that tries to resolve the cosmological lithium discrepancy by means of stellar physics must simultaneously describe these systematic heavy-element trends between dwarfs and giants. Models of atomic diffusion moderated by some additional mixing can indeed explain the observed abundance trends and make quantitative predictions for lithium depletion. The inferred birth-cloud lithium abundances of Spite-plateau stars are higher by $\simeq$0.3\,dex leaving a small ($\simeq$0.15\,dex) nominal offset to the primordial value. 

\keywords{Stars: abundances --
Stars: atmospheres -- Stars: Population II -- Galaxy: globular clusters -- 
Galaxy: abundances -- Cosmology: observations}
}
\maketitle{}

\section{Introduction}

The Spite plateau \citep{Spite82} was originally discovered among field stars as a surprisingly contant value of surface lithium abundances in warm halo stars. Since the 1990s, spectrographs (HIRES, UVES, FLAMES-UVES) at 8-10m telescopes (Keck, VLT) allowed us to probe Spite-plateau stars in globular clusters (\cite{Boesgaard98}, \cite{Gratton01}, \cite{Thevenin01}). As the data quality improved, it became increasingly clear that globular clusters display the same Spite plateau as field stars. 

An even more fundamental question was addressed with GCs around the turn of the century: Are the surface abundances of warm halo stars systematically affected by internal (de)mixing processes like gravitational settling, radiative levitation and rotational mixing? After initial disagreement (\cite{King98}, \cite{Gratton01}, \cite{Thevenin01}), it can now be considered an observational fact that warm halo stars around the turnoff display lower surface abundances than their cooler, more evolved red-giant  counterparts (\cite{Korn06}, \cite{Korn07}, \cite{Lind08}, \cite{Lind09}). 

The physical model most sucessful in explaining this observation is that of atomic diffusion: in the stellar context, the behaviour of inhomogeneous gases (H plus He plus traces of metals) under the influence of gravity and a directed radiation field (\cite{Michaud84}). Under the conditions prevalent in the atmo\-spheres of warm halo stars, both He and metals are expected to slowly sink below the visible layers, as long as processes counteracting this settling (convection, turbulent and/or rotational mixing) are weak. Indeed, the observational trends cannot be explained by atomic-diffusion predictions without introducing a counteracting process that mixes the outer layers more than mixing-length convection does. We refer to it as additional mixing (AddMix with strength T$x.y$ hereafter). 

AddMix is introduced into the stellar-structure models in an ad-hoc manner \citep{Proffitt91} and being a free parameter of the modelling its strength needs to be determined empirically. The range of AddMix efficiencies that keeps the Spite plateau thin and flat is T5.8 to T6.3 \citep{Richard02}. Above T=10$^{6.3}$, significant lithium descruction leads to lower surface lithium abundances at the turnoff, in conflict with observations. 

\section{The cluster sample}

Below, I comment on the analyses of the four globular clusters on which dedicated atomic-diffusion studies have be made.

\begin{table*}
\caption{Inferred, i.e.\ diffusion-corrected, 12-scale lithium abundances in four globular clusters. $\Delta$[Fe/H] refers to the iron trend of the best-fitting Richard model. TOP = turnoff point, RGB = red-giant branch}
\label{abun}
\begin{center}
\begin{tabular}{lcccccccc}
\hline
\\
Cluster& [Fe/H] & $\pm\sigma$ & $\log\varepsilon$(Li) & $\pm\sigma$ & AddMix & $\Delta$[Fe/H] &resolving\\
ref & & & & & T$x.y$ & (TOP\,$-$\,RGB)  & power \\
\\
\hline
\\
{\bf NGC 6397} \\
K06/K07		& $-2.12$ & 0.03 & 2.54 & 0.1 & T6.0 & $-0.14$ & 47k\\
L09         & $-2.10$ & 0.01 & 2.46 & (?) & T6.0 & Fe not studied & 21k \\
N12         & $-2.07$ & 0.03 & 2.57 & 0.1 & T6.0 & $-0.14$ & 47k\\
{\bf NGC 6752} \\ 
G13		& $-1.66$ & 0.04 & 2.58 & 0.1 & T6.2 & $-0.1$ & 47k \\
G14      & $-1.67$ & 0.03 & 2.53 & 0.1 & T6.2 & $-0.1$ & 19k\\
{\bf Messier 4}  \\
M12 & $-1.10$ & 0.07 & (2.7) & (0.1) & (T6.25) & no trend found & 19k\\
N17 & $-1.18$ & 0.02 & 2.59 & 0.1 & T6.2 & $-0.1$ & 19k\\
{\bf Messier 30} \\
G16 & $-2.32$ & 0.03 & 2.48 & 0.1 & T6.0 & $-0.14$ & 19k\\
G16 & $-2.32$ & 0.03 & 2.68 & 0.1 & T5.8 & $-0.23$ & 19k\\
G21 & $-2.3$ & \% & 2.42-2.46 & \% & T6.09-6.2 & $-0.1$ & 23k \\
\\
\hline
\end{tabular}
\end{center}
\end{table*}

\subsection{NGC 6397}

Given its relative proximity, this is by far the best-studied cluster (\cite{Korn06}, \cite{Korn07}, \cite{Lind08}, \cite{Lind09}). It displays {\em element-specific} abundance trends that are well-reproduced by atomic-diffusion models (\cite{Nordlander12}, their Fig. 2). 

A genuine feature of lithium transport in stars was seen for the first time in this cluster: the lithium abundances of stars around 5800\,K are higher by $\simeq$0.1\,dex than those of stars closer to the turnoff. This `lithium hump' signals the onset of the first dredge-up, as lithium is brought from sub-surface layers to the visible surface. Models including atomic diffusion and/or rotational mixing show this effect.

\subsection{NGC 6752}

This cluster is more massive than NGC 6397 and cluster-internal elemental anti-correlations are therefore more pronounced here. Of the ele\-ments studied, this affects Mg the most. In statistical studies of a hundred stars or more, no net effect on the relative abundance {\em trends} is expected. 

Trends in this cluster are weaker than in NGC 6397 and are very similar from element to element, despite only being slightly more metal-rich. There is no sign of a lithium hump. Both these observations ask for more efficient AddMix (\cite{Gruyters13}, \cite{Gruyters14}). 

\subsection{Messier 4}

This cluster was shown not to display any trend in Fe \citep{Mucciarelli11}. However, a more careful analysis (employing an updated temperature scale and NLTE line formation) reveals weak trends for several elements including Fe, even weaker than in NGC 6752 \citep{Nordlander17}. No lithium hump is seen, in agreement with high-AddMix atomic-diffusion models.

This cluster is fairly metal-rich ([Fe/H]\,$\simeq$ $-1$), so spallative production of lithium may have to be considered, if guidance can be taken from field stars. This would lower birth-cloud lithium abundances.

\subsection{Messier 30}

The abundance trends seen in Messier 30 are steeper and more element-specific than in any other cluster studied: Fe shows a trend between turnoff stars and giants of 0.3\,dex, Ca a bimodal trend with the lowest abundances in the middle of the subgiant branch. There is, however, no sign of a lithium hump discernible at the current quality of data (\cite{Gruyters16}, their Figs.\ 5 and 6).

This latter fact leads to an interpretive conflict: Ca and Fe ask for a low efficiency of AddMix, while the absence of a Li hump asks for a moderate-to-high efficiency (see Table 1). If one trusts the heavy elements (which has been the primary strategy), one will invariably infer very high lithium abundances, in full agreement with BBN predictions. 

A fresh look at this cluster with a bluer wavelength setting covering a couple of Mg and Ti lines points towards shallower abundance trends and thus more efficient AddMix values  \citep{Gavel21}. This is ascribed to microturbulence values more appropriate for these metal-poor stars than the ones used by \cite{Gruyters16} (who used the Gaia-ESO microturbulence relation) and a more sophisticated statistical analysis of systematic effects on the abundance trends. The resulting range of most likely AddMix efficiencies (T6.09 - 6.2) gives diffusion-corrected lithium abundances in the range 2.42-2.46. If confirmed by independent studies, this would mean that the lowest-metallicity globular cluster studied in terms of atomic diffusion revealed the lowest diffusion-corrected lithium abundance and could thus point towards a trend of at-birth stellar lithium abundances with metallicity to be corrected for. A necessary extrapolation to Fe/H\,=\,0 would further widen the gap to BBN predictions. Future analyses without parameters like microturbulence and AddMix will surely be able to tell. 

\section{Agreement with BBN?}

Table 1 summarizes the inferred lithium abundances for all studied globular clusters. For a given cluster, the inferred lithium abundances depend heavily on the chosen model, shown here for Messier 30 which with a turnoff apparent magnitude of $m_V$\,=\,18.5 is at the limit of what can be observed with medium-resolution spectrographs at 8m-class telescopes like the VLT. A better understanding of the physics behind AddMix is thus desperately needed, before firm conclusions can be drawn.

With the current set of models, the diffusion-corrected lithium abundances reach levels as high as 
$\log\varepsilon$\,$\simeq$\,2.60 ($n(\mathrm{Li})/n(\mathrm{H})$\,=\,4 10$^{-10}$) falling slightly short of the BBN-predicted value of 2.74 (5.5 10$^{-10}$). However, (dis)agreement at a level corresponding to 1-2$\sigma$ does not warrant a ``cosmological emergency''. The alert level is `orange' at best.      

\section{Implications for the meltdown of the Spite plateau and Galactic archaeology}

Below the metallicity limit of globular clusters ([Fe/H]\,=\,$-2.5$), the field-star Spite plateau changes properties: it seems to partially melt down, that is, more and more stars show surface lithium abundances well below the extra\-polated Spite plateau \citep{Sbordone10}. However, some stars still populate the envelope of the plateau down to the lowest observable metallicities (Gonz\'{a}lez Hern\'{a}ndez, this conference).

As Table 1 indicates, the efficiency of AddMix seem to be a function of metallicity. The lower the metallicity of the cluster, the lower the efficiency of AddMix needed to describe the abundance trends (of Mg, Ca, Ti and Fe or a subset of these) used to infer the preferred model. Given the difficulty of the observations at the lowest metallicities, I would not consider this result as fully established. However, if confirmed by future observations, it could help to explain the partial meltdown of the Spite plateau, as we enter a regime of atomic diffusion with AddMix efficiencies (values below T5.8) which cannot keep the Spite plateau thin an flat (see also Mucciarelli, this conference). In this regime, atomic diffusion is then predicted to have strong 
($>$ 0.3\,dex) element-specific effects on suface abundances, a notion that has to be put to the test using metal-poor binaries. Such binaries have already been analyses and give results not imcompatible with atomic-diffusion predictions (e.g.\ Mel\'{e}ndez, this conference).
  
A marked metallicity dependence of atomic diffusion would also have important implications for Galactic archaeology using non-giant stars, both directly (i.e.\ correcting to the original composition) and indirectly (i.e.\ determining isochrone ages). Higher metallicities and lower ages would be the qualitative result. In this sense, the model grid of \citet{Dotter17} (using a different recipe for AddMix, calibrated on NGC 6397) may only correct atomic-diffusion effects to first order. It may even have unwanted side effects at metallicities higher and lower than that of NGC 6397.

\section{Future work}

This line of research is not finished, far from it! To assess whether or not atomic diffusion contributes to the meltdown of the Spite plateau, fully homogeneous abundance analyses between two or, better, three clusters with a spread in metallicity are needed. 3D+NLTE analyses of Balmer lines and a few spectral lines of key elements have the potential to delineate the trend of T$x.y$ with [Fe/H].

To firmly anchor the absolute lithium abundances, we need to make sure we work on the correct absolute temperature scale. This is favourably done by avoiding issues related to reddening, i.e.\ working purely spectroscopically, and employing effective temperatures of eclipsing binaries near the turnoff as benchmarks. This is a project that would be hard-to-impossible to conduct using field stars emphasizing once again the power of glubular clusters as laboratories.

If it turned out (as current analyses might hint at, see Table 1) that we cannot fully bridge the gap to the predicted BBN lithium abundance, then an additional global (during BBN or later?) or stellar (rotational mixing, lithium depletion during star formation?) process should be considered. 

\section{Conclusions}

Analyses of stars in globular clusters have considerably advanced our understand of the surface abundances of metal-poor stars by overcoming a shortcoming that single field stars inevitably have: we have no knowledge of the composition these singletons were born with. In globular clusters, red giants can provide this knowledge (except for Li, CNO and unobservable He) and reveal that less evolved stars at the main-sequence turn-off and on the subgiant branch show systematically lower heavy-element abundances. Stellar-structure and {-evolution} models treating in detail the physics of atomic diffusion plus some additional mixing below the outer convection zone are capable of describing these trends. These models then predict that the surface lithium abundances of the observed Spite-plateau stars have been lowered as they evolved around the turn-off. Diffusion-corrected birth-cloud lithium abundances reach $\log\varepsilon$\,$\simeq$\,2.60 ($n(\mathrm{Li})/n(\mathrm{H})$\,=\,4 10$^{-10}$) leaving little room for additional effects like non-standard BBN.

The surface abundances of {\em all} warm halo stars are to some degree affected by atomic diffusion. This should be considered in Galactic-archaeology studies of halo stars.  

The latest finding of halo field stars on the hot side of the lithium dip essentially displaying standard-BBN lithium abundances (Lind, this conference) supports the conclusion of the globular-cluster work published in and since 2006: that the cause of the cosmological lithium discrepancy is probably stellar.   
 
\begin{acknowledgements}
As co-chair, I thank all participants of this conference. You made it a success!\\
I am grateful to all the researchers who during the years 2004-2020 have contributed to the ADiOS (Atomic Diffusion in Old Stars) research: in alphabetic order, Martin Asplund, Paul Barklem, Luca Casagrande, Corinne Charbonnel, Remo Collet, Alvin Gavel, Pieter Gruyters, Frank Grundahl, Bengt Gustafsson, Karin Lind, Luydmila Mashonkina, Antonino Milone, Yeisson Osorio, Thomas Nordlander, Nikolai Piskonov, Francesca Primas and, last but not least, Olivier Richard. What would we do without Michaud/Richard models!\\
I acknowledge support from the Swedish National Space Agency (SNSA) and the ChETEC network (COST Action CA16117). 
\end{acknowledgements}

\bibliographystyle{aa}

\end{document}